\documentclass[lettersize,journal]{IEEEtran} 

\usepackage{amssymb,amsmath,graphicx,textcomp, graphicx,indentfirst,float,cite,amsthm,mathrsfs,soul}

\usepackage{graphicx}
\graphicspath{{images/}}

\expandafter\let\csname equation*\endcsname\relax
\expandafter\let\csname endequation*\endcsname\relax
\usepackage{mathtools}
\usepackage{braket}
\usepackage[utf8]{inputenc}
\usepackage[T1]{fontenc}
\usepackage{mathptmx}
\usepackage[markup=underlined]{changes}
\usepackage{hyperref}
\usepackage{cleveref}
\usepackage{etoolbox}
\usepackage{multirow}
\usepackage{orcidlink}
\graphicspath{ {images/} }
\renewcommand{\eqref}[1]{Eq.~(\ref{#1})}

\begin{document}

\title{Modeling Optical Key Distribution over a Satellite-to-Ground Link Under Weak Atmospheric Turbulence}
\author{
    Artur Czerwinski\,\textsuperscript{\orcidlink{0000-0003-0625-8339}}, Miko{\l}aj Lasota\,\textsuperscript{\orcidlink{0000-0002-4038-0330}}, Marcin Jarzyna\,\textsuperscript{\orcidlink{0000-0001-9989-1648}}, 
     Mateusz Kucharczyk\,\textsuperscript{\orcidlink{0009-0003-6679-0447}},    
    Micha{\l} Jachura\,\textsuperscript{\orcidlink{0000-0003-2163-5139}},
     and Konrad Banaszek\,\textsuperscript{\orcidlink{0000-0002-5389-6897}}, \IEEEmembership{Senior Member, IEEE}
     
     \thanks{Received 28 March 2025; revised 9 July 2025; accepted Day Month 2025; date of current version Day Month 2025. This work is part of the project "Quantum Optical Technologies" (FENG.02.01-IP.05-0017/23) carried out within the Measure 2.1 International Research Agendas programme of the Foundation for Polish Science co-financed by the European Union under the European Funds for Smart Economy 2021-2027 (FENG). \textit{(Corresponding author: Artur Czerwinski.)}}
\thanks{Artur Czerwinski was with the Centre for Quantum Optical Technologies, Centre of New Technologies, University of Warsaw, ul. Banacha 2c, 02-097 Warszawa, Poland, and currently is with the Institute of Physics, Faculty of Physics, Astronomy and Informatics, Nicolaus Copernicus University in Torun, ul. Grudziadzka 5, 87-100 Torun, Poland, and also with STARTOVA UMK Sp. z o.o., ul. Gagarina 7, 87-100 Torun, Poland (email: aczerwin@umk.pl)}
\thanks{Miko{\l}aj Lasota is with the Institute of Physics, Faculty of Physics, Astronomy and Informatics, Nicolaus Copernicus University in Torun, ul. Grudziadzka 5, 87-100 Torun, Poland (email: miklas@fizyka.umk.pl).}
\thanks{Marcin Jarzyna is with the Centre for Quantum Optical Technologies, Centre of New Technologies, University of Warsaw, ul. Banacha 2c, 02-097 Warszawa, Poland (email: m.jarzyna@cent.uw.edu.pl).}
\thanks{Mateusz Kucharczyk was with the Centre for Quantum Optical Technologies, Centre of New Technologies, University of Warsaw, ul. Banacha 2c, 02-097 Warszawa, Poland, and currently is with the Faculty of Physics, University of Warsaw, ul. Pasteura 5, 02-093 Warszawa, Poland, and also with Quantum Optical Technologies Sp. z o. o., ul. Lisa Kuli 4, 35-032 Rzesz{\'o}w, Poland (email: mateusz.kucharczyk@qopt.tech).}
\thanks{Micha{\l} Jachura was with the Centre for Quantum Optical Technologies, Centre of New Technologies, University of Warsaw, ul. Banacha 2c, 02-097 Warszawa, Poland, and currently is with the Nokia - Optical Networks, Werinherstraße 91, 81541 Munich, Germany (email: mjachura@nokia.com).} 
\thanks{Konrad Banaszek is with the Centre for Quantum Optical Technologies, Centre of New Technologies, University of Warsaw, ul. Banacha 2c, 02-097 Warszawa, Poland, and also with the Faculty of Physics, University of Warsaw, ul. Pasteura 5, 02-093 Warszawa, Poland (e-mail: k.banaszek@uw.edu.pl).}
}

\maketitle

\begin{abstract}
In this study, we analyze the secret key capacity of intensity modulation/direct detection optical key distribution (IM/DD OKD) for a free-space optical (FSO) link between a low-Earth orbit satellite and an optical ground station. Focusing on downlink communication, we account for atmospheric turbulence, which causes random variations in the transmittance of the FSO channel. We implement an atmospheric channel model that accounts for absorption and scattering, geometric losses, pointing errors, and intensity fluctuations. The secret key capacity is quantified under different noise scenarios and reconciliation code efficiencies, assuming a hard decoding scheme. The performance of the IM/DD OKD protocol is compared under direct and reverse reconciliation regimes. Additionally, we examine the impact of weak and strong wind on the strength of atmospheric turbulence, leading to different results of the secret key capacity. Furthermore, we analyze the characteristics of error distributions that arise from protocol optimization. Our results provide insights into optimizing IM/DD OKD protocols for varying atmospheric conditions.
\end{abstract}


\section{Introduction}

In today's interconnected world, secure communication is crucial for protecting sensitive information exchanged over various networks. Establishing a secret key between two or more parties is one of the possible approaches to achieve secure communication. Among various cryptographic protocols, the recently proposed intensity modulation/direct detection optical key distribution (IM/DD OKD) has received attention due to its practical advantages \cite{ikuta2016intensity,banaszek2021optimization,jarzyna2023quantum}. Compared to the well-established quantum key distribution (QKD) schemes \cite{lo1999unconditional,scarani2009security,pirandola2020advances,xu2020secure}, IM/DD OKD offers significant benefits, such as lower sensitivity to noise and loss, and compatibility with commercially available technology. On the other hand, its level of security can be higher than for the commonly used classical cryptographic protocols.

As wireless systems become more widespread, security arises as a major concern, especially in the upcoming dynamic and decentralized networks. Physical layer security (PLS) has been getting attention as a way to make wireless data exchange more secure \cite{djordjevic2019physical,shakiba2021physical,Abdelsalam2025}. PLS relies on the inherent randomness of wireless channels and unique hardware features to keep data safe. In this context, OKD stands out as a method that fits well with PLS. The security is ensured by the presence of the shot noise that naturally affects the results when someone tries to eavesdrop on an optical signal, making it hard for unauthorized parties to intercept the data \cite{eriksson2018secret,trinh2018design,yamamori2020experimental,trinh2020secrecy}. By using the properties of optical communication systems, OKD provides a reliable way to keep data secret and ensure its integrity. Bringing OKD into the realm of PLS not only strengthens the security of wireless networks, but also shows how optical technology can help tackle the challenges of secure communication in different situations.

Despite its advantages, OKD faces significant challenges in free-space optical (FSO) links due to various atmospheric phenomena, similar to those encountered in other satellite-to-ground optical protocols \cite{liao2017satellite,trinh2019effects,sayat2024satellite,xu2024cooperative,xu2025satellite}. Firstly, the interactions of photons with particles of the atmosphere cause absorption and scattering effects, which reduce the power of the optical signal transmitted to the receiver. For a classical beam, this reduction can be modeled deterministically by the Beer–Lambert law. Secondly, the optical beam is geometrically broadened in the transverse plane during its propagation in free space. Since the distance between a low Earth orbit (LEO) satellite and an optical ground station (OGS) equals a few hundred kilometers, only a small fraction of signal photons can be captured by a receiver telescope, limiting the transmittance of FSO communication channels even further.

In preliminary studies, the IM/DD OKD protocol \cite{banaszek2021optimization} has been applied to estimate the secret key capacity over optical LEO-to-ground links \cite{jachura2023modelling}. However, those studies did not adequately account for the random processes that affect signal transmission. As demonstrated theoretically in Ref.~\cite{banaszek2021optimization}, the performance of the OKD protocol strongly depends on the ratio of transmittance to the eavesdropper and the legitimate receiver. Since this ratio varies due to atmospheric turbulence, it is necessary to evaluate the performance of the protocol under realistic transmission conditions. This work presents a comprehensive analysis of a satellite IM/DD OKD protocol, incorporating two stochastic models of FSO transmission. One of those models describes beam intensity fluctuations caused by random variations in the refractive index along the propagation path. This process follows a log-normal distribution characterized by the scintillation index, which represents the normalized variance of optical intensity \cite{andrews1999theory,andrews2005laser}. Another model accounts for turbulence-induced beam wandering and pointing errors, which appear as deflections of the beam centroid from the receiver’s aperture center. These effects are treated as a random process and can be mathematically modeled by a Weibull distribution \cite{pirandola2021limits}.

This paper is organized as follows: Section \ref{framework} introduces the concept and assumptions of the satellite IM/DD OKD protocol. Section \ref{sec2} describes the atmospheric channel model, while Section \ref{sec3} discusses key generation rates under different reconciliation regimes. Section \ref{sec4} presents the main results, where we compute the secret key capacity---the maximum key amount obtainable during a single satellite pass---under various atmospheric conditions. Finally, Section \ref{secdis} summarizes the significance of our findings for free-space laser communications.

\section{IM/DD OKD protocol}\label{framework}

\begin{figure}[t]
    \centering
    \includegraphics[width=0.725\linewidth]{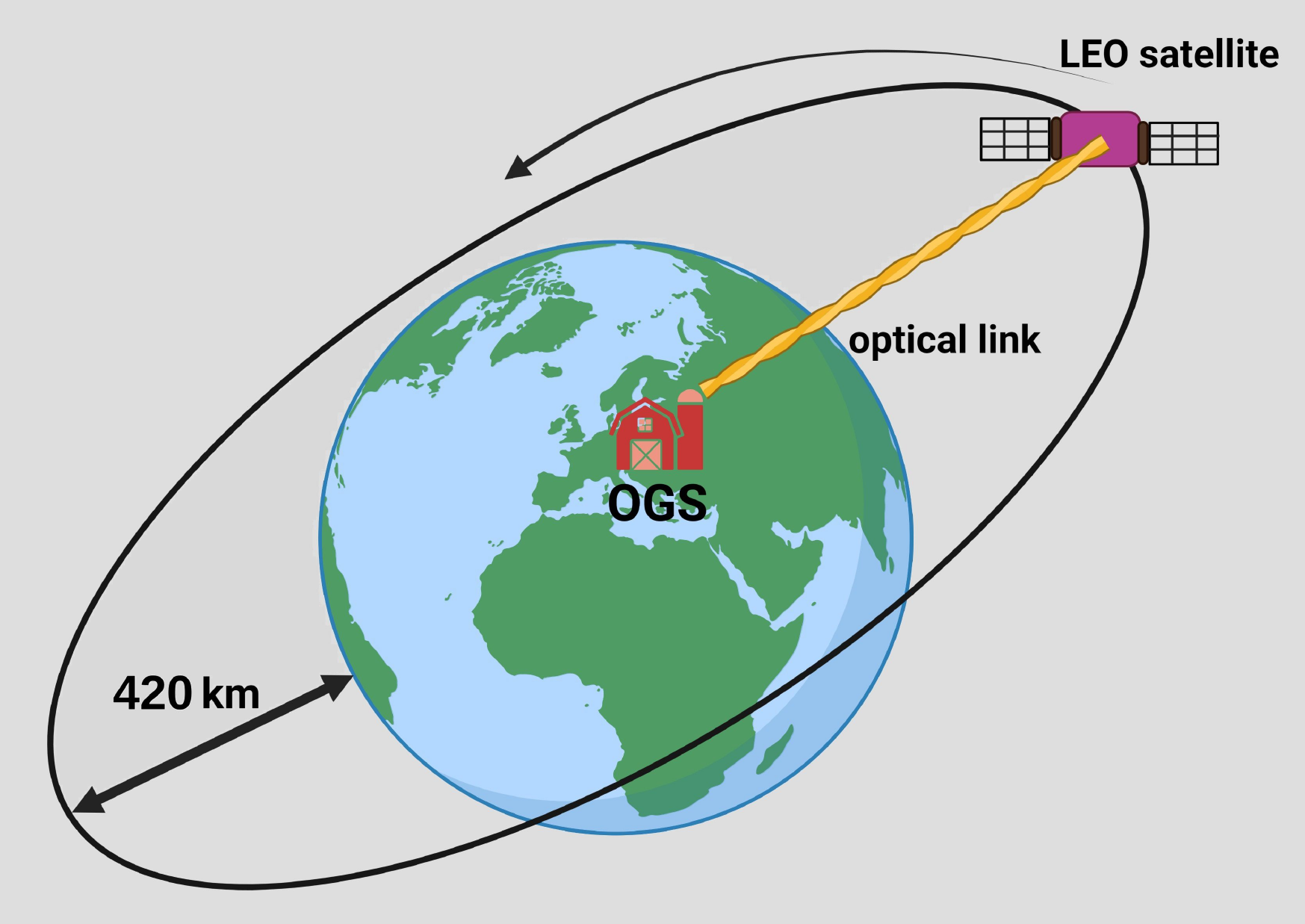}
    \caption{A visualization of a LEO satellite orbiting around the Earth.}
    \label{LEOgraph}
\end{figure}

A simplified scheme of satellite communication over an optical LEO-to-ground link is presented in Fig.~\ref{LEOgraph}. We consider a LEO satellite orbiting the Earth at a constant altitude of 420 km. The orbital parameters match those of the International Space Station. The OGS is located in Toru{\'n}, Poland, at a latitude of $53^\circ$N and an altitude of 65 m above sea level. An optical source mounted on the satellite employs binary intensity modulation for signal encryption \cite{banaszek2021optimization}.

\begin{figure}[t]
    \centering
    \includegraphics[width=0.99\linewidth]{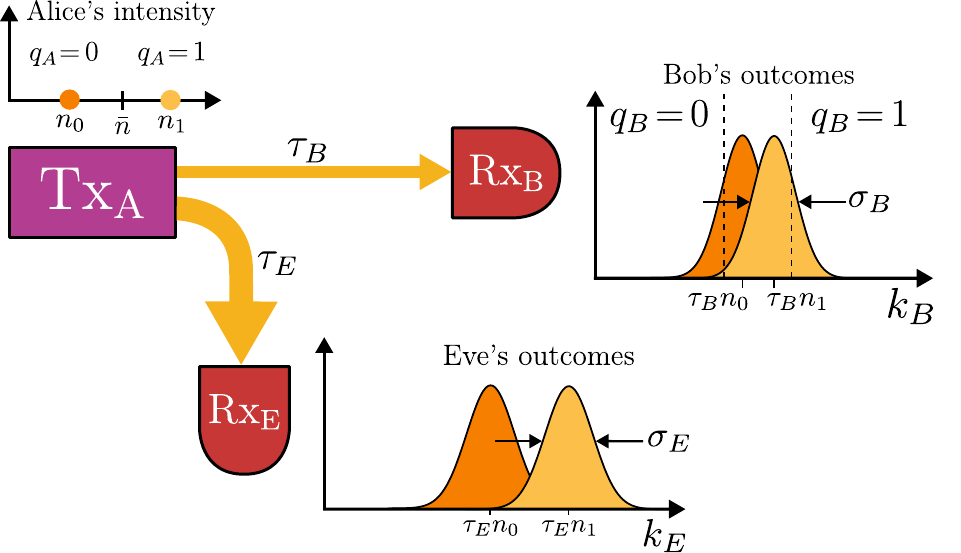}
    \caption{Optical key distribution under passive eavesdropping. The transmitter (Alice) prepares a lightpulse with intensity $n_1$ or $n_0$ representing bit values $1$ and $0$ respectively. The pulses are sent through the atmospheric channel and can be received by the legitimate user (Bob) or an eavesdropper (Eve) attenuated by respective transmission coefficients $\tau_B$ and $\tau_E$, which results in measured intensity values following Gaussian distributions with widths determined by noise variances $\sigma_B^2$ and $\sigma_E^2$ respectively.}
    \label{OKDscheme}
\end{figure}

Fig.~\ref{OKDscheme} shows the IM/DD OKD protocol between two parties, Alice and Bob, under passive eavesdropping. In the binary-modulated scheme, lower and higher intensity levels encode bit values of 0 and 1, respectively. The satellite-based transmitter $\mathrm{Tx_A}$ generates signals with two optical energy levels---lower, denoted by $n_0$, and higher, denoted by $n_1$---corresponding to the bit values $q_A = 0$ and $q_A = 1$, respectively; see the diagram in the upper-left corner of Fig.~\ref{OKDscheme}. Due to inevitable losses and turbulence, only a fraction of the emitted signal reaches the OGS. Simultaneously, another fraction may be intercepted by a passive eavesdropper (Eve) attempting to collect photons and gain access to the information transmitted via the FSO link. As a result, the receiver $\mathrm{Rx_B}$, located at the OGS, receives a fraction of the signal, denoted by $\tau_B$, while $\tau_E$ represents the portion of the signal intercepted by the eavesdropper.

In this model, we assume Gaussian detection noise at each receiving party, with respective variances $\sigma_B^2$ and $\sigma_E^2$, as illustrated conceptually in Fig.~\ref{OKDscheme}. The presence of shot noise ensures the security of the OKD protocol by causing partial overlap in the detected distributions. Consequently, Bob needs to determine threshold values of optical energy (represented by dashed vertical lines in Fig.~\ref{OKDscheme}) to discriminate between the registered bit values $q_B = 0$ and $q_B = 1$.

Following Ref.~\cite{banaszek2021optimization}, we use the modulation depth defined as $\delta_x = \tau_x (n_1 - n_0)/(2 \sigma_x)$ to characterize the measurement results of Bob (for $x=B$) and Eve (for $x=E$). In the hard decoding regime, binary-modulated signals can be discriminated by Bob by defining two symmetric thresholds, $\pm \kappa$, and comparing them with the obtained value of the normalized zero-mean variable $y_B = (k_B - \tau_B \bar{n})/\sigma_B$, where $\bar{n} = (n_0 + n_1)/2$ and $k_B$ denotes the optical energy detected by Bob. Consequently, if the normalized variable $y_B > \kappa$, Bob assigns a bit value of $1$ to the signal, while for $y_B < -\kappa$, Bob assumes the bit value to be $0$.

Precise time synchronization between the satellite transmitter and the OGS receiver is an important practical aspect of the IM/DD OKD protocol. While our model assumes per-symbol hard decision at Bob, we emphasize that accurate synchronization can be achieved using two-way optical ranging techniques, which allow for sub-nanosecond precision in estimating the propagation delay \cite{dirkx2019laser}. Such precision is sufficient to resolve symbol timing with high fidelity, as the typical duration of optical pulses in OKD is on the order of nanoseconds.

Two problems are intrinsically linked to the IM/DD OKD protocol. First, for all received signals such that $-\kappa < y_B < \kappa$, Bob does not discriminate between bit values $0$ and $1$, which implies that this portion of the signal is lost and negatively affects the overall secret key capacity. The second problem relates to the fact that for any threshold $\kappa$, there will always be a nonzero probability of misassigning the bit value to the signal. Both issues can be resolved by optimizing the OKD protocol, which is discussed in detail in Section~\ref{sec3}.


\section{Atmospheric channel modeling}\label{sec2}

In this section, we discuss specific models of various imperfections introduced by the atmospheric channel. We include the effects of both atmospheric absorption and scattering, beam spreading and wandering, pointing errors, and the impact of atmospheric turbulence on overall loss. Modeling the atmospheric channel is crucial for understanding the performance and reliability of OKD in FSO links.

\subsection{Atmospheric absorption and scattering}
\label{Sec:AbsorptionScattering}

The atmospheric absorption and scattering of an optical beam depend on the composition of atmosphere and the local conditions at the moment of optical communication. Absorption depends on the chemical composition of the atmosphere. It can be described by a wavelength-dependent coefficient. Scattering can be divided into Rayleigh scattering and Mie scattering. The former occurs when the optical signal is scattered off of particles with radii much smaller than the optical wavelength of propagating light, while the latter is caused by larger particles of air, \textit{aerosols}, with radii of the order of the wavelength of light. Absorption and scattering together contribute to the wavelength-selective attenuation of the signals \cite{andrews2005laser}.

The Beer-Lambert law is a fundamental model for determining the atmospheric extinction factor $\eta_{\textrm{atm}}$. According to this model
\begin{equation}\label{beerlambert}
    \eta_{\textrm{atm}} = \exp \left(- \int_{0}^{Z} \gamma(h) \, d z \right),
\end{equation}
where $Z$ represents a slant distance travelled by the light and $\gamma(h)$ stands for the altitude-dependent attenuation coefficient that involves both absorption and scattering \cite{bohren2008absorption}. It can be decomposed as
\begin{equation}
    \gamma (h) = \alpha_m (h) + \beta_m (h) + \alpha_a (h) + \beta_a (h),
\end{equation}
where $\alpha_i (h)$ and $\beta_i (h)$ are the molecular ($i=m$) and aerosol ($i=a$) absorption and scattering coefficients, respectively. These factors depend on the transmitted wavelength, the cross-section, and concentrations of the individual particles being absorbed or scattered.

The attenuation coefficient for the atmosphere is commonly modeled as 
\begin{equation}\label{attencoef}
\gamma(h) = \alpha_0 \:\operatorname{exp} ( - h/h_0)
\end{equation}
with the parameters $h_0 = 6 600$ km and $\alpha_0 \approx 5 \cdot 10^{-6}$ m $^{-1}$ corresponding to the sea-level value of the extinction factor for the light at $\lambda = 800$~nm wavelength \cite{vasylyev2019satellite}.

For a satellite that encircles the Earth with the altitude $H$, we introduce a zenith angle $\zeta$ between the vertical direction (zenith) and the pointing direction from the OGS to the satellite. Then, the slant distance $z(h, \zeta)$ between the receiver and an arbitrary point along the propagation path can be computed by using the zenith angle and the altitude of that point \cite{pirandola2021satellite}
\begin{equation}\label{slantdist}
    z(h, \zeta) = \sqrt{h^2 + 2 h R_E + R_E^2 \cos^2 \zeta} - R_E \cos \zeta,
\end{equation}
where $R_E$ denotes the Earth's radius. Similarly, the altitude for a given slant distance can be expressed as
\begin{equation}\label{altitudesat}
     h(z, \zeta) = \sqrt{R_E^2 + z^2 + 2 z R_E \cos \zeta} - R_E.
\end{equation}

Then, the Beer-Lambert law given in Eq.~(\ref{beerlambert}), can be modified to describe FSO communications via arbitrary links. First, let us consider a vertical link for a satellite that is exactly at the zenith, so that its slant distance $z$ is equal to its altitude $H$. Then one obtains
\begin{equation}\label{zenithett}
\eta_{\textrm{atm}}^{\textrm{zen}} = \operatorname{exp} \left( - \int_{0}^{H} \gamma(h) d h   \right) \geq \operatorname{exp} \left( - \alpha_0 h_0 \right) \approx 0.9675,
\end{equation}
which gives a lower bound for the atmospheric channel transmittance, corresponding to a photon loss of $0.143$~dB. In practice, the approximation $\eta_{\textrm{atm}}^{\textrm{zen}} \approx 0.9675$ is valid for any
satellite at the zenith position with the altitude $H \geq 30$~km \cite{pirandola2021satellite}.

In a general situation, the atmospheric extinction factor can be written as
\begin{equation}
      \eta_{\textrm{atm}} (H, \zeta) = \operatorname{exp} \left[ - \alpha_0 \int_0^{z(H, \zeta)} \operatorname{exp}  \left( - \frac{h(z, \zeta)}{h_0} \right)  d z    \right],
\end{equation}
where $z(H, \zeta)$ and $h(z, \zeta)$ can be computed according to Eqs. (\ref{slantdist}) and (\ref{altitudesat}), respectively.

For zenith angles $\zeta \lesssim 1$ rad, the instantaneous height can be approximated by $h(z, \zeta) \approx z  \cos \zeta $ \cite{pirandola2021satellite}. This approximation allows for computing the atmospheric extinction factor \cite{dequal2021feasibility}
\begin{align}\label{finalatmext}
    &\eta_{\textrm{atm}} (H, \zeta) = \operatorname{exp} \left[ - \alpha_0 \int_0^{z(H, \zeta)}  
    \operatorname{exp}  \left( - \frac{z  \cos \zeta}{h_0} \right) d z   \right]  \notag \\
    &= \exp \left\{ \left[-1 + \exp \left(- \frac{z(H, \zeta) \cos \zeta }{h_0}\right)\right]  
    \alpha_0 h_0 \sec  \zeta\right\},
\end{align}
which, after a rough substitution  $\exp \left[- z(H, \zeta) \cos \zeta / h_0\right] \approx 0$, leads to a simplified formula for the effects of atmospheric attenuation
\begin{equation}\label{eq:eta_atm_zen}
\eta_{\textrm{atm}} (\zeta) = \left(\eta_{\textrm{atm}}^{\textrm{zen}} \right)^{\sec \zeta}.
\end{equation}

The above Eq.~(\ref{eq:eta_atm_zen}) shows that atmospheric extinction can be estimated by a one-variable function that depends only on the zenith angle $\zeta$. This equation provides a decent approximation for altitudes greater than $30$~km, becoming almost exact for altitudes beyond $100$~km.

\subsection{Losses due to beam wandering, pointing errors and beam broadening}\label{Sec:PointingErrors}

To model the short-term transmittance associated with turbulence-induced beam wandering and pointing errors, we assume that the Gaussian random walk of the beam centroid around the aperture center of the receiver is well described by the Weibull distribution for the deflection $x$ \cite{dowling1973behavior}, given by the following zero-mean density function \cite{pirandola2021limits}
\begin{equation}
\mathcal{P}_W (x) = \frac{x}{\sigma^2} \operatorname{exp} \left(- \frac{x^2}{2 \sigma^2} \right),
\end{equation}
where the random variable $x$ can take only non-negative values. The centroid of the beam wanders with the total variance of
\begin{equation}\label{totalvar}
    \sigma^2 = \sigma_{\textrm{BW}}^2 + \sigma_{\textrm{p}}^2,
\end{equation}
where $\sigma_{\textrm{BW}}^2$ and $\sigma_{\textrm{p}}^2$ are variances associated with beam wandering and pointing errors, respectively. Short-term transmissivity of the atmospheric channel then reads \cite{vasylyev2012toward}
\begin{equation}\label{etast}
    \eta_{\textrm{st}} (x) = \eta_{\textrm{st}} \,\operatorname{exp} \left[ - \left( \frac{x}{x_0} \right)^{\Gamma} \right],
\end{equation}
where $\eta_{\textrm{st}}$ describes the diffraction-limited damping factor
\begin{equation}
    \eta_{\textrm{st}} = 1 - e^{- 2 a^2_R/ w_{\textrm{st}}^2}
\end{equation}
for a receiver whose aperture is circular with the radius $a_R$ and the short-term beam spot size $w_{\textrm{st}}$. The other symbols in Eq.~(\ref{etast}), $\Gamma$ and $x_0$, denote the shape and scale parameters, respectively, and are given by \cite{vasylyev2012toward}
\begin{align}
 &   \Gamma = \frac{4 \,\eta_{\textrm{st}}^{\textrm{far}} \,\Lambda_1 \left(\eta_{\textrm{st}}^{\textrm{far}}\right)}{1 - \Lambda_0 \left(\eta_{\textrm{st}}^{\textrm{far}}\right)} \left[ \operatorname{ln} \left(\frac{2 \eta_{\textrm{st}}}{1 - \Lambda_0 \left(\eta_{\textrm{st}}^{\textrm{far}}\right)}  \right)\right]^{-1},\\&
x_0 = a_R \left[ \operatorname{ln} \left(\frac{2 \eta_{\textrm{st}}}{1 - \Lambda_0 \left(\eta_{\textrm{st}}^{\textrm{far}}\right)} \right)\right]^{-1/\Gamma}
\end{align}
with $\Lambda_n (x) = \operatorname{exp} (- 2 x) \,I_n(2x)$, where $I_n$ is a modified Bessel function of the first kind with order $n$. Finally, the far-field approximation of the diffraction-limited transmittance, $\eta_{\textrm{st}}^{\textrm{far}}$, takes the form of
\begin{equation}
\eta_{\textrm{st}}^{\textrm{far}} = \frac{2 a_R^2}{w_{\textrm{st}}^2}.
\end{equation}

In downlink communication $\sigma_{\textrm{BW}}^2 \approx 0$, meaning that the beam wandering is caused mainly by the pointing errors (see the details in Appendix). Moreover, the effect of turbulence on the spot size of the beam is also negligible. Therefore, only diffraction affects the beam spreading, which leads to the following approximation
\begin{equation}
    w_{\textrm{st}} \approx w_{\textrm{d}} (z) = w_0 \sqrt{1 + \left(\frac{z}{z_R}\right)^2},
\end{equation}
where $z_R$ is the Rayleigh range and $w_0$ is the initial spot size (the waist radius).

Pointing errors arising from imperfect tracking are assumed to have a standard deviation of $10$~$\mu$rad at the transmitter. Consequently, the pointing error variance, $\sigma_p^2$, is modeled using a simple distance-dependent formula:
\begin{equation}\label{pev}
    \sigma_p^2 = \left(10^{-5} \, z(h, \zeta)\right)^2,
\end{equation}
cf.~\cite{pirandola2021limits}, where $z(h, \zeta)$ denotes the slant distance as in Eq.~(\ref{slantdist}).

\subsection{Atmospheric turbulence-induced loss}
\label{Sec:TurbulenceLoss}

The strength of atmospheric turbulence can be quantified by the refractive index structure parameter, $C_n^2$, which measures the degree of the fluctuations in the refraction index caused by spatial variations of temperature and pressure. This quantity varies with the time of day and geographical location \cite{osborn2018optical}. For a near-ground horizontal link, the value of $C^2_n$ is almost constant. In case of weak turbulence, the refractive index structure parameter is of the order of $10^{-17}$~m$^{-2/3}$, while for strong turbulence it can reach values up to $10^{-13}$~m$^{-2/3}$.

Moving beyond horizontal links, several mathematical models have been proposed to estimate $C^2_n$ as a function of altitude $h$ \cite{andrews2005laser}. A widely used approach is the Hufnagel-Valley (H-V) model, which provides an analytical expression for $C^2_n (h)$. According to this model, the value of $C^2_n (h)$ at an arbitrary altitude is given by \cite{hufnagel1964modulation,valley1980isoplanatic,andrews2005laser,anarthe2023design}:
\begin{align}\label{hvmodel}
    C^2_n (h) &= 8.148 \times 10^{-56} v_{\text{rms}}^2 h^{10} e^{-h/1000} \notag \\
    &\quad + 2.7 \times 10^{-16} e^{-h/1500} + C_0 e^{-h/100}.
\end{align}
 where $C_0$ is the refractive index structure parameter at the sea level (a reference sea-level turbulence value, typically, we assume $1.7 \cdot 10^{-14}$~m$^{-2/3}$ at nighttime and $2.75 \cdot 10^{-14}$~m$^{-2/3}$ at daytime), and $v_{\text{rms}}$ is the root-mean-squared (rms) transverse wind speed in m/s. For the altitude above $5$ km, it can be expressed as
 \begin{equation}
     v_{\text{rms}} = \left( \frac{1}{H_{\textrm{atm}} - 5000} \int_{5000}^{H_{\textrm{atm}}}   V^2(h) dh \right)^{1/2},
 \end{equation}
 where $H_{\textrm{atm}}$ denotes the maximum altitude where the atmosphere exists, $V(h)$ is the vertical altitude-dependent wind profile, and all distance-related quantities are given in meters. This profile is commonly described by the Bufton-Greenwood model \cite{bufton1973comparison}, \cite{greenwood1977bandwidth}, according to which
 \begin{equation}
     V(h) = v_g + v_T \,\operatorname{exp} \left[ - \left( \frac{h  - h_T}{L_T}  \right)^2 \right],
 \end{equation}
 where $v_g$ (m/s) is the ground wind speed, $v_T$ is the wind speed at tropopause, $h_T$ is the altitude of the tropopause, and $L_T$ is the thickness of the tropopause layer. Usually, the last three parameters are fixed with the following values: $v_T = 30$~m/s, $h_T = 12\,448$~m, and $L_T = 4800$~m \cite{roberts2011improved}.

 The variability of the intensity $I$ for an optical beam that propagates through a turbulent atmospheric channel is expressed by the intensity scintillation index (ISI), given by \cite{andrews2000scintillation}
\begin{equation}\label{ISI}
    \sigma^2_I = \frac{\langle I^2 \rangle}{ \langle I \rangle^2} -1,
\end{equation}
which is a normalized variance of the optical intensity at given time and location and the symbol $\langle \cdot \rangle$ represents the expected value. Because intensity fluctuations are an ergodic process in time and space, the normalized variance can be calculated by averaging the spatial and temporal intensity statistics \cite{giggenbach2008fading}. In weak-fluctuation regime, where the scintillation index in Eq.~(\ref{ISI}) is less than unity, the derived expressions for the scintillation index indicate that it is proportional to the Rytov variance \cite{andrews2000scintillation}
\begin{equation}
    \sigma_1^2 = 1.23 \, C_n^2 \, k^{7/6}\, L^{11/6},
\end{equation}
where $ k = 2 \pi/\lambda$ is the wavenumber and $L$ is the length of the propagation path between the transmitter and the receiver. For the weak-turbulence regime the Rytov variance physically represents the irradiance fluctuations associated with an unbounded plane wave. Otherwise, it is considered a measure of the optical turbulence strength when extended to strong-fluctuation regimes by increasing either $C_n^2$ or the path length $L$, or both.

In the weak-turbulence approximation, a more accurate theoretical formula for evaluating the ISI of a plane wave is the Rytov index, $\sigma^2_R$. It can be computed based on the value of $C^2_n$ along the propagation path \cite{kaushal2016optical}:
\begin{equation}\label{rytovindex} \sigma^2_I \approx \sigma^2_R = 2.25 k^{7/6}  (\sec \zeta)^{11/6} \int_{h_g}^H C^2_n (z) (z - h_g)^{5/6} \:d z, \end{equation}
where $h_g$ is the altitude of the OGS above sea level; $\zeta$ and $H$ denote the zenith angle and the altitude of the transmitter, respectively, as described in Sec. \ref{Sec:AbsorptionScattering}. The ISI, approximated by Eq.~(\ref{rytovindex}), models intensity fluctuations caused by atmospheric turbulence, which are characterized by the random variable $\mathscr{L}$. This variable follows a log-normal distribution with the probability density function given by \cite{andrews2005laser,trinh2022statistical}:
\begin{equation} p_{\textrm{int}} (\mathscr{L}) = \frac{1}{ \mathscr{L} \sqrt{2 \pi \sigma^2_I}} \exp{\left( - \frac{\left( \ln \mathscr{L} + \sigma^2_I/2 \right)^2}{2 \sigma^2_I} \right)}. \end{equation}

The expected value of the random variable $\mathscr{L}$ is $\mathbb{E} [\mathscr{L}] = 1$, which means that, over longer timescales, the average impact of atmospheric turbulence on signal intensity is neutral \cite{alquwaiee2016asymptotic}.

\subsection{Overall atmospheric channel transmittance}

To summarize, the channel transmittance for a LEO satellite-to-ground link depends on three main degradation factors, discussed in Sec.\,\ref{Sec:AbsorptionScattering}--\ref{Sec:TurbulenceLoss}: deterministic loss due to attenuation and scattering $ \eta_{\textrm{atm}} (\zeta)$, geometric loss caused by beam spreading and pointing errors  $\eta_{\textrm{st}} (x)$, and the turbulence-induced effect due to intensity fluctuations $\mathscr{L}$. The contributions from two of these factors, namely $\eta_{\textrm{st}} (x)$ and $\mathscr{L}$, are random and statistically independent, as supported by the channel models in Refs.~\cite{ntanos2021leo,pirandola2021limits,pirandola2021satellite,trinh2022statistical}. Moreover, we also need to include an internal loss related to the efficiency of the setup, which is denoted by $\eta_{int}$. As a result, the overall channel transmittance, $\eta$ can be calculated as
\begin{equation}\label{transmittance1}
    \eta = \eta_{\textrm{int}} \,\eta_{\textrm{atm}} (\zeta) \, \eta_{\textrm{st}} (x)\, \mathscr{L}.
\end{equation}

For a channel with fluctuating transmittance, given by Eq.~(\ref{transmittance1}), there is always a non-zero probability for an instantaneous $\eta$ to be larger than unity. To resolve this problem some studies have proposed closed-form expressions for the probability density function of transmittance, vanishing for $\eta > 1$ \cite{vasylyev2012toward,vasylyev2016atmospheric,vasylyev2018theory}. However, for a typical LEO satellite-to-ground communication channel the overall losses are considerably high, implying that even for the case of large turbulence-induced intensity fluctuations the resulting value of $\eta$ falls within the physical limit of $\eta \in [0,1]$ \cite{trinh2022statistical}.

The formula for channel transmittance, Eq.~(\ref{transmittance1}), could in principle be extended by including an additional factor to account for angle-of-arrival (AoA) fluctuations, see more in Refs.~ \cite{dabiri2021uav,maharjan2022atmospheric,xu2023outage}. AoA fluctuations can introduce additional fading in the received optical signal, thereby degrading system performance and increasing the outage probability. However, consistent with prior works \cite{ntanos2021leo,pirandola2021limits,pirandola2021satellite,trinh2022statistical}, we assume that this effect can be mitigated by accurate tracking and alignment systems at the ground station. Such systems help maintain efficient coupling and reduce the impact of AoA-induced fading \cite{trinh2023experimental}.

\subsection{Coherence time and orbital slicing}\label{SECcoherencetime}

Coherence time, $\tau_0$, indicates for how long one can assume that the atmospheric conditions remain roughly stable. It can be estimated from the formula \cite{farley2022reference}
\begin{equation}\label{coherencetime}
    \tau_0 =  \left(2.91 \: k^2 \, \int_0^{\infty} C^2_n (h) [V(h)]^{5/3} d h \right)^{- 3/5},
\end{equation}
where we again use the H-V model for $C^2_n (h)$ and the Bufton-Greenwood wind model for $V(h)$. Under the weak turbulence regime, Eq.~(\ref{coherencetime}) gives results belonging to the range of $5-7$~ms, depending on specific values of the parameters. Therefore, it is justified to assume that the transmittance through the air is fixed for periods of the order of $1$~ms. Although by studying the strength of the atmospheric turbulence in the time domain, one can see that the received intensity of light is not constant during the coherence time, the occurring intensity fluctuations are insignificant enough to support this assumption. 

To confirm the above assumption we generate a time series from a Gaussian process using a log-normal distribution with the Kolmogorov model for the scintillation index. First, we define the log-amplitude as \cite{ishimaru1978wave}
\begin{equation}
    \chi := \frac{1}{2} \operatorname{ln} \frac{I}{I_0},
 \end{equation}
 where $I_0$ denotes the average received intensity. Since the intensity is assumed to follow the log-normal distribution with mean value $I_0$, the log-amplitude can be modeled by a normal distribution with zero mean and the variance of \cite{ishimaru1978wave,lyras2020scintillation}
\begin{equation}\label{logamvar}
    \sigma_{\chi}^2 = (2 \pi)^2 \int_{0}^L d h \int_{0}^{\infty} d \varkappa \, H_{\textrm{pl}}^2(L - h, \varkappa) \Phi (\varkappa, h) \varkappa,
\end{equation}
where $L$ is the slant path through the turbulent medium,
\begin{equation}
    H_{\textrm{pl}}^2(L - h, \varkappa)  = k \operatorname{sin} \left( \frac{L - h}{2 k} \varkappa^2 \right)
\end{equation}
and $\Phi (\varkappa, h)$ denotes the Kolmogorov spectrum
\begin{equation}
\Phi (\varkappa, h)  = 0.033\, C_n^2 (h) \varkappa^{-11/3}.
\end{equation}
The argument $\varkappa$, can be interpreted as the magnitude of the spatial frequency vector, expressed in the units of rad/m.

\begin{figure}[t]
    \centering
\includegraphics[width=0.875\linewidth]{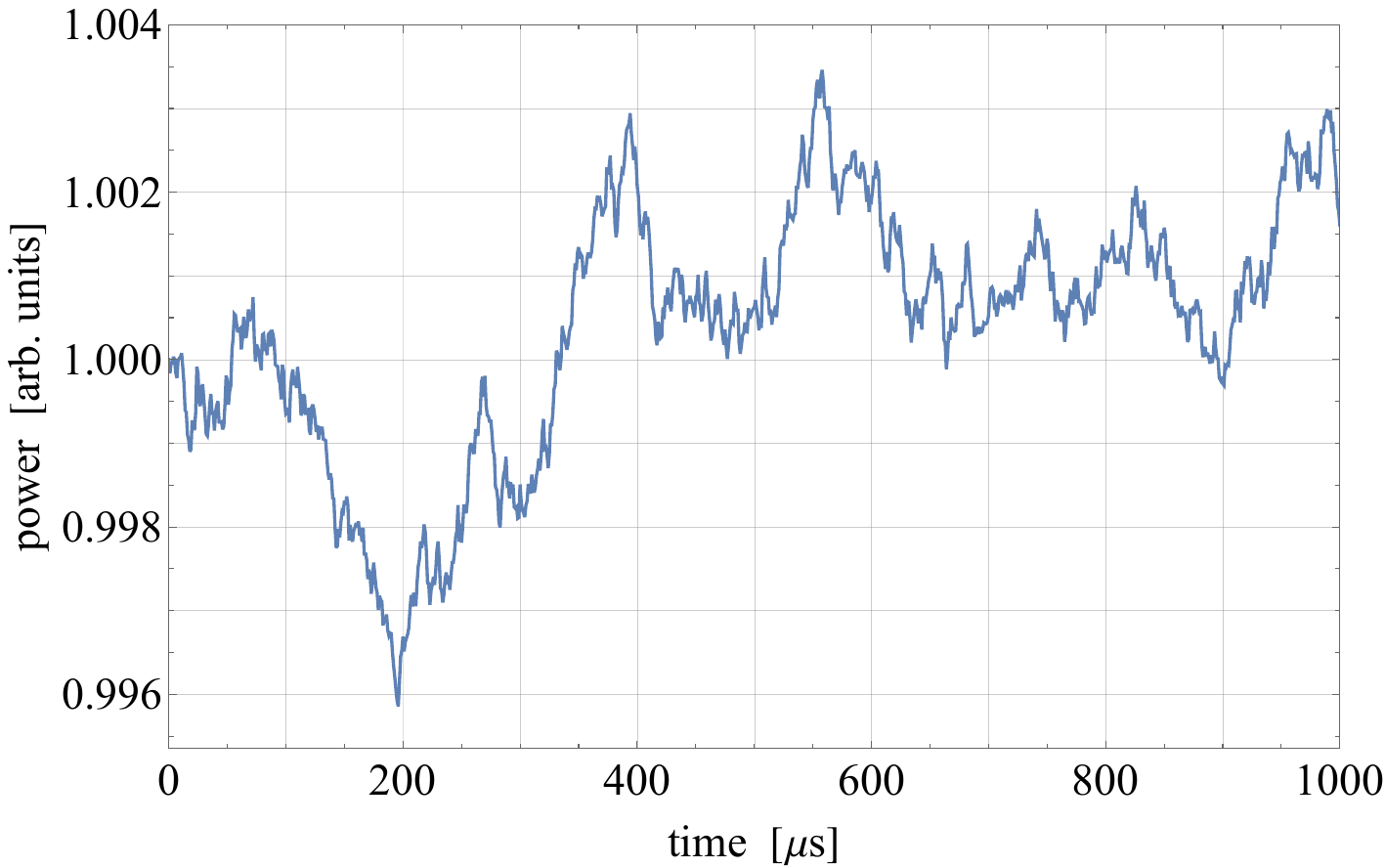}
    \caption{Simulated time series of the received optical power for $1$~ms assuming a weak turbulence model with clear sky and nighttime conditions with average power $I_0=1$ and source positioned at $420$~km altitude and zenith angle $55^\circ$.}
    \label{simplot}
\end{figure}

Scintillation time series are generated via stochastic differential equations driven by fractional Brownian motion (fBm) \cite{mishura2008stochastic}. For a given variance, $\chi (t)$ is a Gaussian process with zero mean, which implies that one can use the fractional Langevin equation to simulate the time series of log-amplitude
\begin{equation}
    d \chi(t) = - \Lambda \chi (t) d t + \sigma d B_{\mathcal{H}},
\end{equation}
where $d B_H$ denotes the increments of the fBm with the Hurst index $\mathcal{H}$ \cite{frisch1995turbulence,benassi2000identification}. The parameters $\Lambda$ and $\sigma$ depend on the dynamic properties of the stochastic process \cite{lyras2020scintillation}. 

By determining these parameters and computing the log-amplitude variance according to Eq.~(\ref{logamvar}), we can follow the received power in the time domain, as presented in Fig.~\ref{simplot}. The average power was set arbitrarily as $I_0 = 1$. The results were obtained assuming a weak-turbulence model with clear sky and nighttime conditions. The position of the source was fixed in the sky with an altitude equal to $420$~km and a zenith angle $55^{\circ}$. A typical wavelength for FSO communications was assumed, i.e., $\lambda = 1\,550$~nm. From Fig.~\ref{simplot} one can conclude that the variability of the optical power due to atmospheric turbulence is insignificant over the period of $1$~ms since the received signal changes by less than $0.5\%$. Therefore, we can discretize a satellite pass over the OGS by dividing the transit time of the satellite into $1$ms intervals and fixing the transmittance of the atmospheric channel for each of them. This procedure is called orbital slicing. This approach allows us to compute the key amount independently in every time slot and later calculate the sum over the entire satellite pass.

\subsection{Satellite motion and communication time}

We assume that a LEO satellite encircles the Earth in the same direction as the Earth rotates, \emph{i.e.} from west to east. Then, its angular velocity relative to the OGS can be expressed as
\begin{equation}
    \omega_r  = \sqrt{\frac{G M_E}{(R_E + H)^3}}- \frac{2 \pi}{T_E},
\end{equation}
where $G$ is the gravitational constant, while $M_E$ and $T_E$ denote the Earth's mass and rotation period, respectively. The transit time of the satellite between its zenithal position ($\zeta = 0$) and an arbitrary location on its path around the sky ($\zeta = \theta <\pi/2$) can be computed as \cite{pirandola2021satellite}
\begin{equation}
    t(\theta,H) = \omega_r^{-1} \operatorname{arccos} \left(\frac{R_E + z(H, \theta) \operatorname{cos} \theta}{R_E + H} \right).
\end{equation}

Optical communications can be performed only within a limited range of the zenith angle so that we avoid critical photon loss and turbulence that occur for low elevations. We assume that the optimal sector for communication exists within one radian from the zenith. Therefore, the effective time for communication reads
\begin{equation}
    t_{\textrm{eff}} = 2  \, t(1,H).
\end{equation}

Based on the analysis outlined in Sec.~\ref{SECcoherencetime}, we divide the effective time into $1-$ms intervals, assuming that within each interval the atmospheric conditions remain unchanged. 

\section{Key generation rates}\label{sec3}

The performance of the IM/DD OKD protocol is analyzed under two reconciliation regimes: direct and reverse. These regimes differ in terms of the key generation rate, which represents the amount of key attainable per slot. In both cases, however, the key generation rate is given by an analytical formula that can be maximized to evaluate the optimal performance of the protocol. The distinction between direct and reverse reconciliation regimes is analogous to reconciliation strategies used in continuous-variable QKD protocols \cite{grosshans2003quantum,laudenbach2018continuous,pan2020secret}. In both types of protocols Bob first communicates to Alice in which time bins he obtained a meaningful bit value (i.e. $0$ or $1$) without revealing the measured value and then Alice leaves only the corresponding bits in her sequence. The difference between the two approaches appears in terms of who performs the error correction of the remaining bit sequences. In direct reconciliation this task is performed by Bob, whereas in the reverse reconciliation it is Alice who needs to correct her sequence.

\subsection{Direct reconciliation regime}\label{SECdirect}

In the direct reconciliation approach, Alice first determines which bits in her sequence correspond to meaningful ones measured by Bob. She then applies a predefined error-correcting code to compute a syndrome, which she sends to Bob. This syndrome provides additional information that allows Bob to correct errors in his sequence, ensuring it matches Alice's. Under this protocol, in the hard decoding regime, the key generation rate is given by:
\begin{equation}\label{keyamount}
    \mathcal{K}_d = \mathrm{max} \,\{ p_{\textrm{raw}} \left[ 1 - f \, h(\epsilon) - I(A:E) \right], 0  \}.
\end{equation}
where $p_{\textrm{raw}}$ denotes the probability of generating a raw key bit, $f \geq 1$ accounts for limited reconciliation efficiency related to the classical protocols of error correction and privacy amplification performed during the postprocessing stage, $h(\epsilon) := - \epsilon \log_2 \epsilon - (1- \epsilon) \log_2 (1 - \epsilon)$ is the binary entropy for the error probability $\epsilon$, and $I(A:E)$ stands for the mutual information between the sender and the eavesdropper. The quantities $p_{\textrm{raw}}$ and $\epsilon$ can be expressed in terms of Bob's modulation depth $\delta_B$ and threshold value $\kappa$ as \cite{banaszek2021optimization}
      \begin{equation}\label{prawep}
      \begin{aligned}
        &   p_{\textrm{raw}} = \frac{1}{2} \left[ \mathrm{erfc}\left( \frac{\kappa + \delta_B}{\sqrt{2}}\right) + \mathrm{erfc}\left( \frac{\kappa - \delta_B}{\sqrt{2}}\right)\right], \\ &
         \epsilon = \frac{1}{2 p_{\textrm{raw}}} \,\mathrm{erfc}\left( \frac{\kappa + \delta_B}{\sqrt{2}}\right),
      \end{aligned}
    \end{equation}
where $\mathrm{erfc}$ denotes the complementary error function.

The mutual information $I(A:E)$ specifies how much information about the signals transmitted by Alice can be obtained by Eve. It reads \cite{kunz2020low}
\begin{equation}
     I(A:E) = \sum_{q_A} \int_{- \infty}^{+ \infty} \, p_A (q_A) p_{E|A} (y_E|q_A) \log_2 \frac{p_{E|A} (y_E|q_A)}{p_{E} (y_E)} d y_E,
     \label{eq:IAE}
\end{equation}
where $p_A (q_A)$ is the probability for Alice to send bit value $q_A$, equal to 1/2 for $q_A=0,1$ in the considered case, 
\begin{equation}
    p_E (y_E) = \frac{1}{2 \sqrt{2 \pi}} \left[ \exp \left(- \frac{(y_E + \delta_E)^2}{2}\right) + \exp \left(- \frac{(y_E - \delta_E)^2}{2}\right) \right]
\end{equation}
is the probability for Eve to obtain the normalized value of $y_E$ during the signal intensity measurement and
\begin{equation}
    p_{E|A} (y_E|q_A) = \begin{cases} (\sqrt{2 \pi})^{-1} \,\exp \left[ - (y_E + \delta_E)^2/2 \right] & q_A = 0 \\
    \\ (\sqrt{2 \pi})^{-1} \exp \left[ - (y_E - \delta_E)^2/2 \right] & q_A =1. \end{cases}
\end{equation}
is the probability for Eve to obtain $y_E$ conditioned on Alice sending the bit value $q_A$. Eq.\,(\ref{eq:IAE}) can be transformed into
\begin{equation}
    I(A:E) = \delta^2_E \log_2 e - \int_{- \infty}^{+ \infty} \, p_{E} (y_E) \log_2 \left[ \cosh \left(\delta_E y_E\right)\right] \,d y_E.
\end{equation}

\subsection{Reverse reconciliation regime}

In the reverse reconciliation approach, in contrast to the direct reconciliation protocol, it is Bob who calculates the syndrome and sends it to Alice, who then performs error correction on her sequence. The formula for key generation rate reads then 
\begin{equation}\label{keyamount2}
    \mathcal{K}_r = \mathrm{max} \,\{ p_{\textrm{raw}} \left[ 1 - f \, h(\epsilon) - I(B:E) \right], 0  \},
\end{equation}
where $p_{raw}$ and $\epsilon$ are given by Eq.~(\ref{prawep}), and $I(B:E)$ denotes the mutual information between Bob and Eve, which can be computed as \cite{banaszek2021optimization}
\begin{equation}
\begin{aligned}
  &  I(B:E) = \\& \int_{-\infty}^{\infty} \frac{dt}{\sqrt{2 \pi}} \exp\left[ - (t - \delta_E)^2/2\right] \left\{ \epsilon \log_2 \left[ \epsilon e^{\delta_E t} + (1 - \epsilon) e^{- \delta_E t} \right]\right.  \\&
   \left.+(1 - \epsilon) \log_2 \left[ \epsilon e^{- \delta_E t} + (1 - \epsilon) e^{\delta_E t} \right] - \log_2 [\cosh{(\delta_E t)}] \right\}.
\end{aligned}
\end{equation}

\subsection{Assumptions of the noise model}

The formulas Eqs. (\ref{keyamount}) and (\ref{keyamount2}) for key generation rate depend on several factors, including the modulation depths $\delta_E$ and $\delta_B$. These two parameters are connected by the relation \cite{banaszek2021optimization}
\begin{equation}\label{scaling}
    \delta_B =  \frac{\delta_E}{\sqrt{\mathscr{E}}},
\end{equation}
where the symbol $\mathscr{E}$ is referred to as "eavesdropper advantage" and is defined as
\begin{equation}\label{ratio}
    \mathscr{E} =  \left( \frac{\tau_E \, \sigma_B}{ \tau_B \, \sigma_E} \right)^2.
\end{equation}
In this work, we consider two specific cases of $\mathscr{E}$. In an idealized scenario of Bob's and Eve's detectors operating at the shot noise level, we obtain $\sigma_E^2 = \tau_E \bar{n}$ and $\sigma_B^2 = \tau_B \bar{n}$, and consequently,
\begin{equation}\label{Rshot}
    \mathscr{E} = \frac{\tau_E}{\tau_B},
\end{equation}
which means that the eavesdropper advantage depends solely on the relative transmittance between the Bob's and Eve's channels. On the other hand, in a more realistic scenario, Bob's and Eve's detection systems can be affected by both the shot noise and the thermal noise. In such a case, the variance of the detector reads $\sigma_x^2 = \tau_x \bar{n} + \sigma_{x,\textrm{th}}^2$ (for $x = B$ or $x = E$). The latter element, i.e. $\sigma_{x,\textrm{th}}^2$, provides the contribution due to the thermal noise. If one now assumes that both detectors feature the thermal noise, the expression for $\mathscr{E}$ takes the form 
\begin{equation}\label{Rthermal}
\mathscr{E} =  \frac{\tau_E(1+\xi_B)}{\tau_B(1+\xi_E)},
\end{equation}
where the quotient $\xi_x=\sigma_{x,\textrm{th}}^2/(\tau_x \bar{n})$ (for $x=B,E$) can be interpreted as the strength of Bob's or Eve's detector thermal noise in relation to the shot noise.

\section{Simulation results - the secret key capacity}\label{sec4}

The goal of this section is to estimate the secret key capacity, which represents the maximum amount of secret key that can be generated using the IM/DD OKD protocol during one zenithal satellite pass in a downlink communication affected by atmospheric turbulence. To determine the secret key capacity, we divide the total communication time into $1-$ms intervals where atmospheric conditions remain approximately constant. Using the key generation rate formulas from Section~\ref{sec3}, we optimize the performance of the protocol by selecting discrimination thresholds based on the atmospheric conditions. Finally, we compute the secret key capacity by summing the contributions over the entire satellite pass.

\begin{table}[t!]
\centering
\caption{Parameters and assumptions used in numerical simulations of the seceret key capacity.}
\label{parameters}
\begin{tabular}{l c}
\hline
\textbf{Atmospheric Turbulence} & \\
\hline
Refractive index model & Hufnagel–Valley \\
$C_n^2$ at the sea level ($C_0$) & $1.7 \cdot 10^{-14}$~m$^{-2/3}$ \\
Coherence time & 1 ms \\
Weak (strong) wind  ($ v_{\text{rms}}$) & 21~m/s (57~m/s)\\
\hline
\textbf{Transmitter Parameters} & \\
\hline
Wavelength & $1550$~nm \\
Symbol duration & $1$~ns \\
Initial beam waist & $1$~cm \\
\hline
\textbf{Receiver Parameters} & \\
\hline
Telescope radius (OGS) & $15$~cm \\
Telescope radius (Eavesdropper) & $20$~cm / $30$~cm \\
Altitude (OGS / Eavesdropper) & $65$~m / $65$~m \\
Internal loss factor ($\eta_{\text{int}}$) & 0.5 \\
\hline
\textbf{Satellite Properties} & \\
\hline
Orbit apogee/perigee & $420$~km / $420$~km \\
Comm. session length & $166$~s \\
\hline
\end{tabular}
\end{table}

\subsection{Relative transmittance for weak and strong wind}

We consider a zenithal pass of a LEO satellite orbiting the Earth in a circular trajectory at an altitude of $H = 420$~km. In this case the effective communication time is equal to $t_{\textrm{eff}} = 166$~s, which is divided into $1-$ms intervals. For each slot we simulate an instantaneous atmospheric transmittance according to Eq.~(\ref{transmittance1}). As for detection efficiency, the extinction factor accounting for this loss can be approximated as $\eta_{\textrm{int}} \approx 0.5$, corresponding to photon loss of $3$~dB \cite{liorni2019satellite}. We focus on the scenario with Eve having a larger aperture telescope and experiencing no pointing errors or random fluctuations of signal intensity during her measurement process, which is advantageous for her. The specific values of the parameters characterizing the system and the atmospheric channel used in numerical simulations are collected in Table \ref{parameters},

\begin{figure}[t!]
   \centering
\includegraphics[width=0.9\columnwidth]{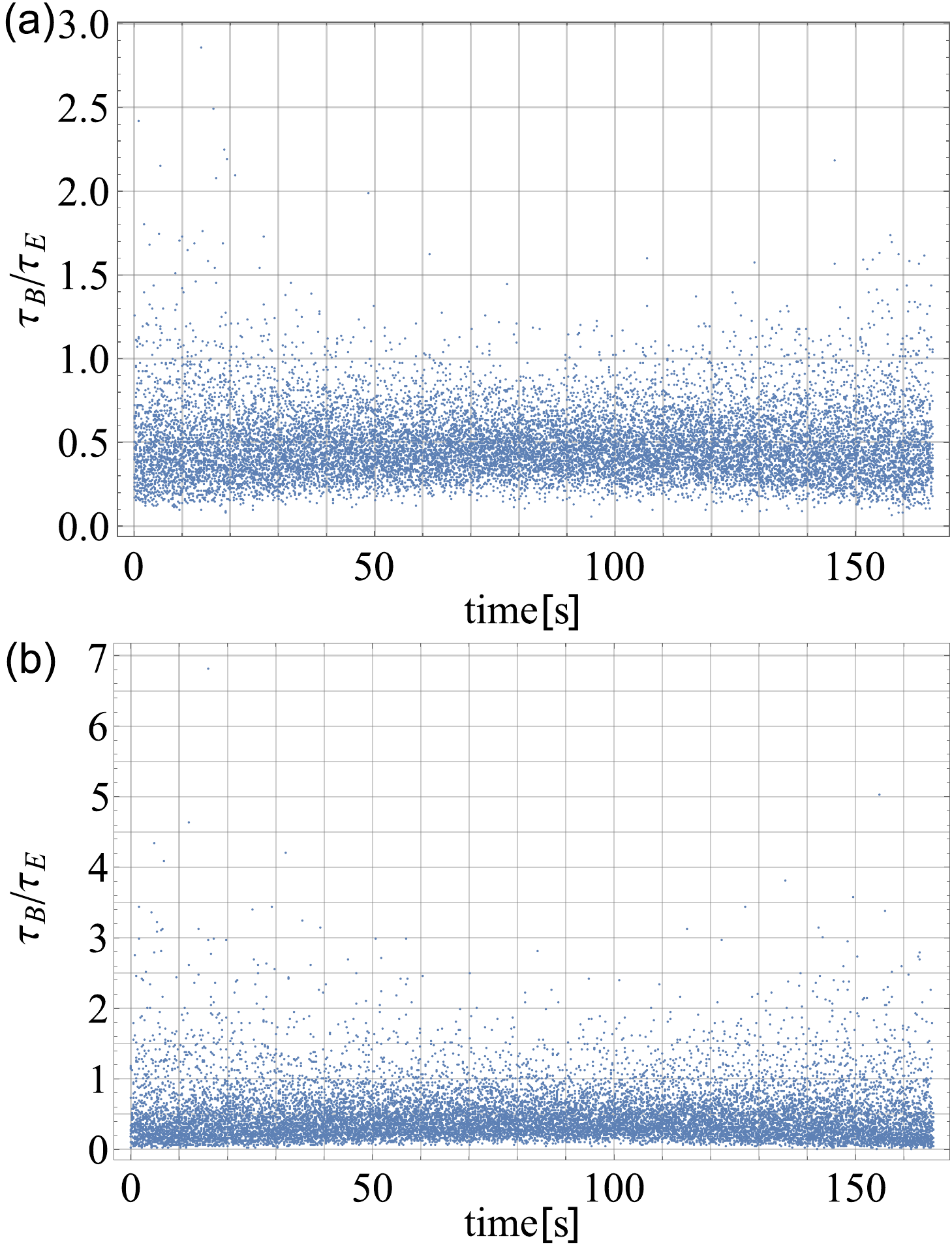}
\caption{Relative transmittance $\tau_B/\tau_E$ between the recipient (Bob) and eavesdropper (Eve) for a zenithal pass of a LEO satellite during (a) weak wind and (b) strong wind under assumptions from Table~\ref{parameters} and radii for Bob's and Eve's telescopes aperture given by $a_B = 15$~cm and $a_E = 20$~cm respectively.}
\label{relativetransmittance}
\end{figure}

Since the transmittance of the atmospheric channel connecting Alice and Bob is affected by turbulence, the relative transmittance $\tau_B/\tau_E$ varies in time, leading to different key generation rates in corresponding $1-$ms time intervals. The simulated values of $\tau_B/\tau_E$ for the satellite pass, restricted to the length of the communication session, are presented in Fig.~\ref{relativetransmittance}, assuming $a_B = 15$~cm (radius of Bob's telescope) and $a_E = 20$~cm (analogous parameter for Eve). For comparison, we juxtapose results for weak wind ($ v_{\text{rms}} = 21$~m/s) in Fig.~\ref{relativetransmittance}(a) and strong wind ($ v_{\text{rms}} = 57$~m/s) in Fig.~\ref{relativetransmittance}(b). Due to intensity fluctuations, the ratio $\tau_B/\tau_E$ randomly exceeds $1$. However, we do not intend to benefit from random leaps of transmittance. Thus, in further calculations, all the cases of $\tau_B/\tau_E > 1$ shall be replaced by $\tau_B/\tau_E = 1$, which is a conservative assumption.

The values of $\tau_B/\tau_E$ can be used to compute the key generation rate according to Eq.~(\ref{keyamount}) or (\ref{keyamount2}), while $\mathscr{E}$ is given either by Eq.~(\ref{Rshot}) (communication limited by the shot noise only) or Eq.~(\ref{Rthermal}) (thermal noise included). In the latter scenario, we assume that the strength of the detector thermal noise is either approximately equal to the shot noise ($\xi_x = 1$) or two times stronger ($\xi_x = 2$). Different combinations concerning Bob's and Eve's detectors are considered.

For each $1-$ms time interval, when $\mathscr{E}$ is fixed, we optimize the formula for the key generation rate by determining such a discrimination threshold $\kappa$ that maximizes the key. In the final step, we add the key amounts calculated for separate time intervals over the entire timespan of the communication session, which results in the secret key capacity.

\subsection{Results for direct reconciliation}

\begin{table}[t]
\centering
\caption{Results of the secret key capacity for a zenithal pass of a LEO satellite. Assumptions: direct reconciliation under weak and strong wind conditions.}
\renewcommand{\arraystretch}{1.1}
\begin{tabular}{lcccc}
\hline
 &\multicolumn{4}{c}{Secret key capacity} \\
 \hline
Rec. factor & $f=1.0$  & $f=1.1$  & $f=1.2$  & $f=1.3$  \\
\hline
\multicolumn{5}{c}{Weak wind conditions} \\
\hline
\multicolumn{1}{c}{$a_E = 20$ cm} \\ 
$\xi_B = 0, \xi_E = 0$ & 7.41 Gb & 6.71 Gb & 6.16 Gb & 5.73 Gb \\ 
$\xi_B = 1, \xi_E = 0$ & 1.52 Gb & 1.28 Gb & 1.10 Gb & 970 Mb \\ 
$\xi_B = 2, \xi_E = 0$ & 430 Mb  & 336 Mb  & 274 Mb  & 231 Mb \\ 
$\xi_B = 2, \xi_E = 1$ & 3.17 Gb & 2.76 Gb & 2.45 Gb & 2.22 Gb \\ 
\multicolumn{1}{c}{$a_E = 30$ cm} \\ 
$\xi_B = 0, \xi_E = 0$ & 1.09 Gb & 893 Mb & 758 Mb & 659 Mb \\ 
$\xi_B = 1, \xi_E = 0$ & 86.9 Mb & 61.9 Mb & 47.3 Mb & 37.7 Mb \\ 
\hline
\multicolumn{5}{c}{Strong wind conditions} \\
\hline
\multicolumn{1}{c}{$a_E = 20$ cm} \\ 
$\xi_B = 0, \xi_E = 0$ & 7.03 Gb & 6.43 Gb & 5.95 Gb & 5.57 Gb \\ 
$\xi_B = 1, \xi_E = 0$ & 1.44 Gb & 1.27 Gb & 1.03 Gb & 919 Mb \\ 
$\xi_B = 2, \xi_E = 0$ & 371 Mb & 290 Mb & 236 Mb & 200 Mb \\ 
$\xi_B = 2, \xi_E = 1$ & 2.87 Gb & 2.59 Gb & 2.37 Gb & 2.14 Gb \\ 
\multicolumn{1}{c}{$a_E = 30$ cm} \\ 
$\xi_B = 0, \xi_E = 0$ & 1.03 Gb & 856 Mb & 732 Mb & 641 Mb \\ 
$\xi_B = 1, \xi_E = 0$ & 82.3 Mb & 58.6 Mb & 44.8 Mb & 34.2 Mb \\ 
\hline
\end{tabular}
\label{combined_table_direct}
\end{table}

The results of the secret key capacity are summarized in Table \ref{combined_table_direct}, presenting findings under conditions of weak and strong wind, respectively. The calculations were conducted considering four reconciliation factors. Notably, $f = 1.0$ represents an ideal scenario for reconciliation efficiency, albeit unattainable in practice, serving as a benchmark for comparison. The subsequent cases, where $f = 1.1$, $f = 1.2$, and $f = 1.3$, reflect less favorable scenarios, necessitating more raw key bits for error correction and privacy amplification. As anticipated, the secret key capacity diminishes with an increase in the parameter $f$, indicating a reduction in reconciliation efficiency.

Furthermore, it is evident that the inclusion of thermal noise in the analysis leads to a significant decrease in the secret key capacity. Specifically, when considering $\xi_B = 2$ and $\xi_E = 0$, characterizing a case in which Bob experiences thermal noise twice as strong as shot noise while Eve operates at the shot noise level, the secret key capacity decreases by over an order of magnitude compared to a scenario when both Bob and Eve are constrained solely by the shot noise. Incorporating thermal noise in Eve's setup serves to balance the conditions, resulting in an overall increase in the total key amount.

Importantly, increasing Eve's telescope aperture to $a_E = 30$~cm noticeably deteriorates the generated key, as demonstrated under both the shot noise limit and a scenario involving thermal noise. Notably, in the case characterized by parameters $\xi_B = 1$, $\xi_E = 0$, and $a_E = 30$~cm, when only Bob is affected by thermal noise and Eve possesses a telescope twice as large as Bob's, the secret key capacity is significantly reduced.

Comparing the results in Table \ref{combined_table_direct} with respect to the wind conditions, it becomes evident how the assumption of stronger wind impacts the key amount. Stronger wind conditions result in a higher scintillation index, thereby amplifying the effects of atmospheric turbulence, which subsequently reduces the secret key capacity. This highlights the importance of accounting for atmospheric conditions, particularly wind strength, in satellite optical key distribution systems.

\subsection{Results for reverse reconciliation}

\begin{table}[t]
\centering
\caption{Results of the secret key capacity for a zenithal pass of a LEO satellite. Assumptions: reverse reconciliation under weak and strong wind conditions.}
\renewcommand{\arraystretch}{1.1}
\begin{tabular}{lcccc}
\hline
 & \multicolumn{4}{c}{Secret key capacity} \\
 \hline
Rec. factor & $f = 1.0$ & $f = 1.1$ & $f = 1.2$ & $f = 1.3$ \\
\hline
\multicolumn{5}{c}{Weak wind conditions} \\
\hline
\multicolumn{1}{c}{$a_E = 20$ cm} \\ 
$\xi_B = 0, \xi_E = 0$ & 18.8 Gb & 15.1 Gb & 11.6 Gb & 8.83 Gb \\
$\xi_B = 1, \xi_E = 0$ & 3.85 Gb & 2.90 Gb & 2.0 Gb  & 1.50 Gb \\
$\xi_B = 2, \xi_E = 0$ & 1.06 Gb & 760 Mb & 515 Mb  & 356 Mb \\
$\xi_B = 2, \xi_E = 1$ & 7.93 Gb & 6.24 Gb & 4.61 Gb & 3.40 Gb \\
\multicolumn{1}{c}{$a_E = 30$ cm} \\
$\xi_B = 0, \xi_E = 0$ & 2.69 Gb & 2.02 Gb & 1.43 Gb & 1.02 Gb \\
$\xi_B = 1, \xi_E = 0$ & 214 Mb  & 140 Mb  & 86 Mb   & 58 Mb   \\
\hline
\multicolumn{5}{c}{Strong wind conditions} \\
\hline
\multicolumn{1}{c}{$a_E = 20$ cm} \\
$\xi_B = 0, \xi_E = 0$ & 17.2 Gb & 13.4 Gb & 10.3 Gb & 7.82 Gb \\
$\xi_B = 1, \xi_E = 0$ & 3.46 Gb & 2.61 Gb & 1.77 Gb & 1.22 Gb \\
$\xi_B = 2, \xi_E = 0$ & 976 Mb  & 731 Mb  & 504 Mb  & 348 Mb  \\
$\xi_B = 2, \xi_E = 1$ & 7.27 Gb & 5.21 Gb & 3.87 Gb & 2.62 Gb \\
\multicolumn{1}{c}{$a_E = 30$ cm} \\
$\xi_B = 0, \xi_E = 0$ & 2.37 Gb & 1.70 Gb & 1.26 Gb & 874 Mb \\
$\xi_B = 1, \xi_E = 0$ & 193 Mb  & 136 Mb  & 77.4 Mb & 52.0 Mb \\
\hline
\end{tabular}
\label{combined_table_reverse}
\end{table}

The investigation extended to including the reverse reconciliation approach reveals additional insights into the capabilities of a satellite IM/DD OKD protocol, see Table \ref{combined_table_reverse}. Despite retaining the previously observed trends, such as the impact of noise, telescope sizes, and wind strength on key generation, a notable observation emerges: the secret key capacity achieved through reverse reconciliation tends to surpass that of direct reconciliation for specific parameter configurations. This suggests that while both models exhibit similar responses to varying system parameters, reverse reconciliation offers a more favorable outcome in terms of the secret key capacity under specific conditions.

In comparing direct and reverse reconciliation regimes, it is essential to discuss their fundamental differences and potential implications. Direct reconciliation involves processing the raw key bits to remove errors and amplify privacy directly at the receiver's end. Conversely, reverse reconciliation shifts the error correction burden, and thus also the computational load, to the sender. Importantly, in reverse reconciliation, less information is revealed about the transmitted bit sequence, which reduces the effective power of the eavesdropper and improves the secret key capacity, as demonstrated in Table \ref{combined_table_reverse}.

\subsection{Analysis of error distributions}

\begin{figure}[t]
     \centering
            \includegraphics[width=0.9\columnwidth]{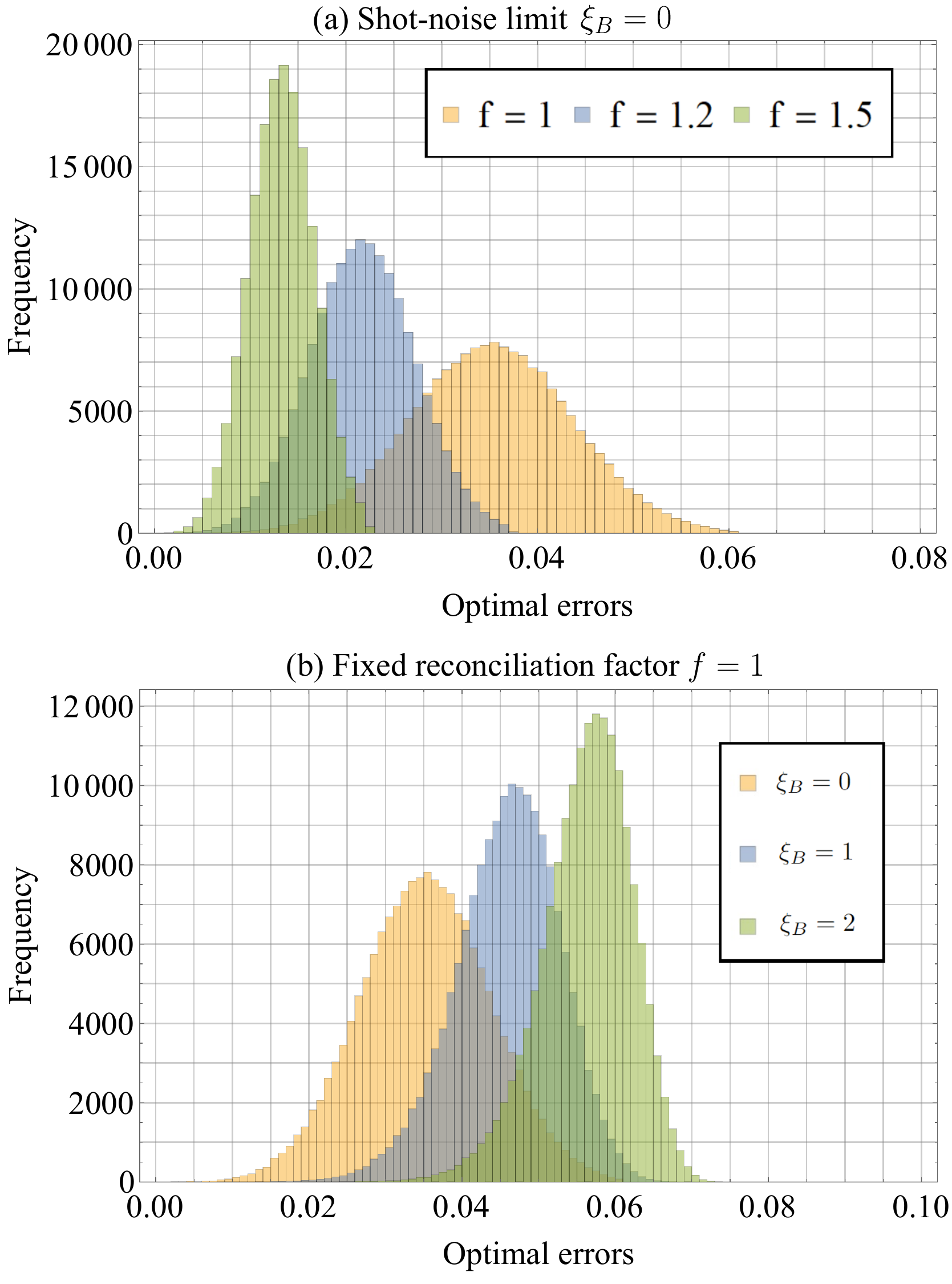}
            \caption{Histograms presenting distributions of optimal errors determined by maximizing the key amount under parameters gathered in Table~\ref{relativetransmittance} and eavesdropper telescope aperture radius $a_E=20$~cm.}\label{thregraphs}
\end{figure}

Let us again consider the formula for the key generation rate in the direct reconciliation scheme, i.e. Eq.~(\ref{keyamount}). The error rate, $\epsilon$, depends on the discrimination threshold $\kappa$ through the bottom formula in Eq.\,(\ref{prawep}). Thus, every adjustment of $\kappa$ performed for a particular $1-$ms time interval causes some change of $\epsilon$. As a result, for a satellite pass, we obtain a distribution of optimal errors that correspond to every $1-$ms interval. The error rate histograms obtained for the weak wind are presented in Fig.~\ref{thregraphs}, assuming parameters as in Table \ref{parameters} with $a_E = 20$~cm. For clarity, we selected only three values of reconciliation factors to illustrate how it affects the error distribution under the shot-noise limit, see Fig.~\ref{thregraphs}(a). On the other hand, in Fig.~\ref{thregraphs}(b), we demonstrate the impact of thermal noise on the error distribution, assuming perfect reconciliation (i.e., $f=1$).

Our analysis reveals two notable phenomena related to error distributions in OKD. First, by observing Fig.~\ref{thregraphs}(a), we see that when the reconciliation factor increases, the error distribution becomes narrower and shifts to the left, which implies that lower error rates are expected to maximize the key rate. This phenomenon highlights the importance of suitable codes for error correction, which can allow the system to operate effectively with lower error rates. Secondly, from Fig.~\ref{thregraphs}(b), we notice that the introduction of thermal noise narrows the error distribution and shifts it to the right, signifying a preference for higher error values when operating under noisy conditions.
Furthermore, as the strength of the thermal noise increases, this shift becomes more significant. This observation presents the trade-off between maximizing the key generation rate and tolerating higher error rates in the presence of environmental noise.

\subsection{Benchmarking of the results}

Benchmarking the results of this study against those reported in the literature, particularly in the context of satellite-based QKD, is not straightforward. Most existing works present the secret key rate (in bits/s), typically as a function of the zenith angle. In contrast, we intentionally focus on the total secret key attainable during a single satellite pass, i.e., the secret key capacity. We argue that this cumulative measure is more relevant for practical satellite missions than the instantaneous key rate.

Direct comparisons are further complicated by substantial differences in the underlying assumptions used across studies, including the specific QKD protocol employed, atmospheric channel models, and satellite characteristics. Therefore, any reported key rate or capacity should be interpreted with caution.

Nevertheless, it is instructive to reference a recent study on satellite QKD in which a secret key is generated between two users via a trusted satellite in low-Earth orbit, with communication to ground stations along the satellite’s path. In that work, the asymptotic key limit per pass was reported to reach up to 4.58 Mbit \cite{roger2023real}. From Tab.~\ref{combined_table_direct}, we see that for $\xi_B = 1$, $\xi_E = 0$ and $a_E = 30$ cm, the IM/DD OKD protocol yields a higher secret key capacity, indicating the potential advantage of this scheme in realistic free-space channels.

Moreover, the QuantSat-PT case study modeled a LEO satellite at an altitude of 750 km and reported that, under optimal conditions (i.e., the satellite at zenith), a maximum key rate of $32.1$ kbit/s could be achieved \cite{galetsky2022leo}. Assuming a 5-minute pass duration, this corresponds to approximately $9.6$ Mbits of sifted key, with the final secret key being somewhat lower after error correction and privacy amplification. This example reinforces the general expectation that, over a typical satellite pass lasting a few minutes, satellite QKD can generate a total secret key length on the order of several megabits.

Experimentally, such secret key lengths were confirmed by the Chinese Micius mission---a dedicated quantum science satellite launched from Jiuquan, China, to an altitude of 500 km \cite{yin2017satellite}. In 2024, it was reported that the Eurasian-scale experimental satellite-based QKD performed with Micius achieved an overall sifted key of 2.5 Mbits during a single communication session \cite{khmelev2024eurasian}.

These current limits in satellite QKD illustrate that the IM/DD OKD protocol may offer a significant advantage, as our model reports secret key capacities on the order of several gigabits under ideal noise conditions (i.e., the shot-noise limit). However, as shown in Tables~\ref{combined_table_direct} and \ref{combined_table_reverse}, the achievable values are highly sensitive to thermal noise and eavesdropper advantage. Therefore, the benefits of the IM/DD OKD protocol can be maintained only if low-noise detectors are available.

\section{Discussion}\label{secdis}

The results obtained from our research offer practical implications for the development of secure and efficient implementations of the IM/DD OKD protocol. These findings provide a foundation for tailoring key distribution systems to specific requirements, for example optimizing for higher key generation rates in the presence of thermal noise.

Importantly, our analysis deliberately assumes a scenario that is advantageous to the eavesdropper, with Eve having access to telescopes larger than those used by the legitimate receiver Bob. Despite this, the IM/DD OKD protocol demonstrates robust performance, maintaining high secret key capacity under these challenging conditions. This indicates that the protocol can effectively generate secure keys even when the adversary has certain practical advantages. Such resilience highlights the potential of the OKD approach as a practical alternative or complement to conventional IM/DD QKD schemes, particularly in free-space satellite-to-ground links where equipment asymmetries can occur.

In the realm of satellite OKD, the comparative analysis between direct and reverse reconciliation regimes accentuates the practical considerations surrounding computational burden and system architecture. While reverse reconciliation provides higher theoretical secret key capacities under the investigated parameter configurations, the feasibility of its implementation is questioned due to the inherent limitations in computational power aboard the satellite, which serves as the sender in this scheme. In contrast, direct reconciliation, which places the computational burden on the recipient (i.e., the OGS) with access to relatively large computational resources, emerges as a more practical choice. Despite potentially yielding a lower secret key capacity, the scalability and operational feasibility of direct reconciliation make it a preferred option for a satellite OKD systems, emphasizing the importance of aligning reconciliation strategies with the computational capabilities of the system components.

Furthermore, our study highlights the critical influence of atmospheric conditions on key distribution rates. By comparing scenarios with weak and strong wind, we observed significant variations in the transmittance of the atmospheric channel, affecting instantaneous key generation rates. Under strong wind conditions, the increased turbulence and resulting fluctuations in signal intensity adversely affect the secret key capacity, demonstrating the necessity of accounting for atmospheric dynamics in system design. This comparison provides valuable insights into optimizing key distribution protocols for varying environmental conditions, ultimately enhancing the reliability and security of satellite-based OKD systems.

Finally, the analysis of error distributions in the binary-modulated OKD protocol offers valuable insights about the trade-offs between key generation, error rates, thermal noise, and reconciliation efficiency. These findings highlight the importance of evaluating different reconciliation methods to optimize secret key capacity.

This study contributes to ongoing efforts to develop secure and efficient cryptographic systems while providing a foundation for further research. Future studies could examine how aperture averaging mitigates scintillation effects and improves secret key capacity \cite{carrillo2025statistics}. Additionally, it would be worthwhile to investigate the potential performance gains of soft decoding schemes in free-space implementations of the OKD protocol.

Finally, an interesting direction for future development involves the integration of the IM/DD OKD protocol with satellite-based QKD systems. While QKD ensures information-theoretic security, it is often constrained by stringent loss and hardware requirements. In contrast, the IM/DD OKD protocol offers a more flexible and potentially robust approach under less ideal conditions, particularly in high-loss regimes. A hybrid deployment—where QKD is used during optimal atmospheric windows or in high-priority communication scenarios, and IM/DD OKD serves as a fallback or parallel secure channel—could enhance the overall availability and resilience of satellite-to-ground secure links. Such a strategy could be particularly beneficial for global-scale quantum communication networks with heterogeneous hardware and variable link quality.

\section*{Data availability}
The data supporting the results reported in this paper are available upon reasonable request from aczerwin@umk.pl


{\appendix[Turbulence-Induced Beam Wander in Downlink Channels]

Turbulence-induced beam wander refers to the random displacement of the centroid of an optical beam caused by large-scale inhomogeneities in the refractive index of the atmosphere. These inhomogeneities arise due to turbulence and act as a series of weak lenses that slowly deflect the beam’s path during propagation. As a result, the position of the beam spot fluctuates at the receiver plane \cite{churnside1990wander}.

In the context of satellite-based optical downlinks, turbulence-induced beam wander primarily occurs in the lower part of the atmosphere, close to the receiver. However, due to the geometric propagation from high altitudes and the relatively short optical path within turbulent layers, its overall effect is generally small compared to pointing errors or scintillation.

The variance of the beam centroid displacement caused by turbulence, denoted by $\sigma_{\text{BW}}^2$ in Eq.~(\ref{totalvar}), often called beam wander variance, can be estimated using the following formula for a collimated Gaussian beam \cite{andrews2005laser}:
\begin{equation}\label{bwv}
    \sigma_{\text{BW}}^2 = 2.42\, C_n^2\, L^3\, w_0^{-1/3},
\end{equation}
where $L$ is the path length through the turbulent medium, and $w_0$ is the beam waist at the transmitter.

In this simplified model, $C_n^2$ is treated as a constant to provide an upper-bound estimate of beam wander. One can interpret this constant as an effective path-averaged value for the refractive index fluctuations along the propagation direction.

\begin{figure}[t]
    \centering
    \includegraphics[width=0.9\linewidth]{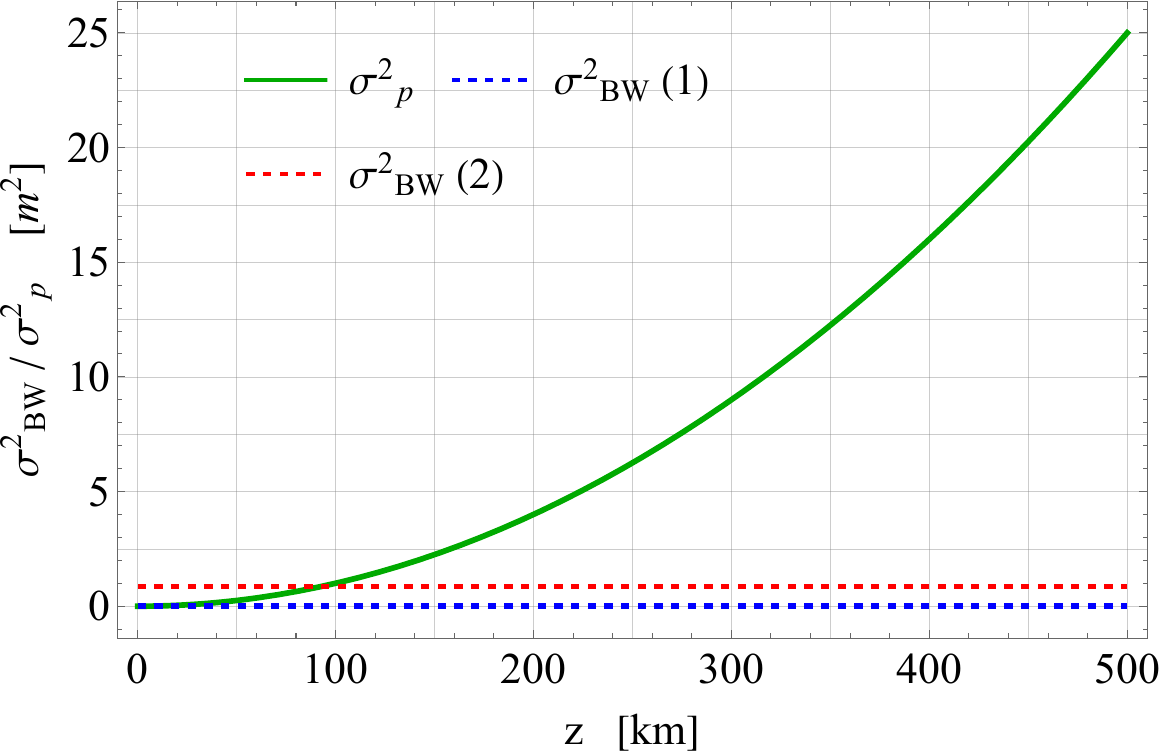}
    \caption{Pointing error variance, $\sigma_p$, as a function of propagation distance $z$, compared with two values of the beam wander variance $\sigma_{\text{BW}}$, where (1) corresponds to $C_n^2 = 4 \cdot 10^{-16}$ m$^{-2/3}$ and (2) represents $C_n^2 = 4 \cdot 10^{-14}$ m$^{-2/3}$.}
    \label{sigmaBWplot}
\end{figure}

In our simulation, we assumed that the impact of $\sigma_{\text{BW}}^2$ can be neglected, as this quantity is small compared to $\sigma_p^2$. To support this assumption, we present a plot of $\sigma_p$ (as defined in Eq.~(\ref{pev})) along with $\sigma_{\text{BW}}^2$ (from Eq.~(\ref{bwv})). For the beam wander variance, we assume a propagation length of $L = 12\,448$ m, which corresponds to the tropopause height at mid-latitudes. We consider two values of $C_n^2$: one representing very weak turbulence, $C_n^2 = 4 \cdot 10^{-16}$ m$^{-2/3}$, and another reflecting a more realistic turbulence scenario, $C_n^2 = 4 \cdot 10^{-14}$ m$^{-2/3}$.

The results are shown in Fig.~\ref{sigmaBWplot}. We observe that for the very weak turbulence case (1), $\sigma_{\text{BW}}^2$ is negligible and close to zero. In the more realistic scenario (2), we obtain $\sigma_{\text{BW}}^2 \approx 0.87$ m$^2$. Comparing $\sigma_{\text{BW}}^2$ (2) with $\sigma_p^2$, we find that for propagation distances up to $z = 93$ km, $\sigma_{\text{BW}}^2 > \sigma_p^2$, indicating that beam wander would dominate the total variance in Eq.~(\ref{totalvar}). However, since $\sigma_p^2$ increases with the propagation path, for a LEO satellite at an altitude of $H = 420$ km, we find $\sigma_p^2 = 17.84$ m$^2$, which clearly justifies our assumption to neglect $\sigma_{\text{BW}}^2$.

Although the contribution of $\sigma_{\text{BW}}^2$ is insignificant for LEO satellite scenarios, this effect may become relevant in future studies involving shorter vertical links, such as inter-city FSO communication, and should be taken into account accordingly.}

\bibliography{Ref}
\bibliographystyle{IEEEtran}

\begin{IEEEbiographynophoto}{Artur Czerwinski}
received his Ph.D. degree in Physics from Nicolaus Copernicus University in Toru\'{n}, Poland, in 2018, specializing in Mathematical and Quantum Physics under the supervision of Prof. Andrzej Jamio{\l}kowski. During his Ph.D. studies, he also conducted research at the Center for Theoretical Physics of the Polish Academy of Sciences. From 2019 to 2022, he worked at the Institute of Physics, Nicolaus Copernicus University in Toru\'{n}, contributing to the project {\it Applications of Single-Photon Technologies}, led by Prof. Piotr Kolenderski. He subsequently held a postdoctoral position at the Centre of New Technologies, University of Warsaw (2022–2023), working on a project in quantum optical technologies. Currently, he is an Assistant Professor at Nicolaus Copernicus University in Toru\'{n}, working in the Single Photon Applications Laboratory. At the same time, he serves as a Lead Researcher at Startova.pl sp. z o.o., a special purpose vehicle established for technology deployment. He is actively involved in projects funded by the European Space Agency, including {\it CCSDS Standardized Ranging for Optical Communication Terminals}, {\it Multi-dimensional Free-space Quantum Key Distribution Protocol}, and {\it Space-Based Links for the Quantum Internet}.
\end{IEEEbiographynophoto}

\begin{IEEEbiographynophoto}{Miko{\l}aj Lasota}
received his MSc and PhD degrees from Nicolaus Copernicus University in Torun in 2010 and 2014, respectively. Following his doctoral studies, he worked as a postdoctoral researcher at Palack\'{y} University in Olomouc, Czechia. Since 2017, he has been an Assistant Professor at Nicolaus Copernicus University in Toru\'{n}. He leads a grant funded by the National Science Centre in Poland and contributes to ESA-funded projects.
\end{IEEEbiographynophoto}

\begin{IEEEbiographynophoto}{Marcin Jarzyna}
Marcin Jarzyna received the M.Sc. and Ph.D. degrees in physics from the University of Warsaw, Warsaw, Poland, in 2011 and 2016 respectively, where he was mainly focused on quantum metrology. He has been with the Centre of New Technologies, University of Warsaw from 2016 to 2023 and then he held a senior researcher position at the Department of Optics, University of Olomouc, Czechia in 2024. Currently, he is at the Centre for Quantum Optical Technologies, Centre of New Technologies, University of Warsaw since 2025. His current research interests include optical communication in low power limit, deep-space and satellite optical communication, quantum key distribution and impact of signal amplification and quantum signal processing in quantum and conventional optical technologies.
\end{IEEEbiographynophoto}

\begin{IEEEbiographynophoto}{Mateusz Kucharczyk}
received his M.Sc. degree in Physics from the University of Warsaw, Poland, in 2024, specializing in methods for secret key distillation in optical key distribution. His master’s thesis, supervised by Prof. Konrad Banaszek and Dr. Michał Jachura, was awarded 2nd prize in the 2024 Polish Electronics Society competition for the best optoelectronics thesis. He presented his work at the ECOC 2024 conference and co-authored a conference paper at OFC 2024. Currently, he is pursuing a Ph.D. at the Faculty of Physics, University of Warsaw, while also working at Quantum Optical Technologies sp. z o.o., a technological enterprise, as part of an industrial Ph.D. program. His research interests include optical communication, physical layer security, error correction codes, and passive optical networks.
\end{IEEEbiographynophoto}

\begin{IEEEbiographynophoto}{Micha{\l} Jachura}
received his M.Sc. and Ph.D. degrees in physics from the University of Warsaw, Warsaw, Poland, in 2014 and 2018, respectively. During his doctoral studies, he served as an Academic Visitor at the University of Oxford, Universität Paderborn, and the University of St. Andrews. After obtaining his Ph.D., he worked as an R\&D Optical Engineer for several technology companies, including 3Shape, Solvemed Inc., and Quantum Optical Technologies sp. z o.o., as well as a Senior Scientist at the Centre for Quantum Optical Technologies, University of Warsaw. In 2024, he joined Coriant R\&D GmbH (currently a part of Nokia), where he works as an Optical Line Systems Architect. He is an avid fan of Allan Holdsworth's music.
\end{IEEEbiographynophoto}

\begin{IEEEbiographynophoto}{Konrad Banaszek}
received the M.Sc. and Ph.D.
degrees in physics from the University of Warsaw, Warsaw, Poland, in 1997
and 2000, respectively. He held postdoctoral positions with the University of
Rochester, NY, USA, and the University of Oxford, Oxford, U.K., followed
by a Junior Research Fellowship with St. John’s College, Oxford, U.K., and
faculty appointments with the Nicolaus Copernicus University, Toru\'{n}, Poland,
from 2005 to 2009 with the University of Warsaw, since 2009.
He was the first director of the Centre for Quantum Optical Technologies
established in 2018 by the University of Warsaw in partnership with the University
of Oxford under the International Research Agendas Programme operated
by the Foundation for Polish Science. His research interests include quantum
physics and optical sciences with a focus on novel approaches to communication,
sensing, and imaging that enable operation beyond the standard quantum limits.
Dr. Banaszek is a corresponding Member of the Polish Academy of Sciences
and Fellow of Optica (formerly OSA). He was an Associate Editor for Optics
Express. In 2001, he received the European Physical Society Fresnel
Prize for his contributions to the understanding of non-classical light and its
applications in quantum information processing.
\end{IEEEbiographynophoto}
\end{document}